
\magnification\magstep1
\scrollmode

\input xy       
\xyoption{all}  

\font\eightrm=cmr8                
\font\tensmc=cmcsc10              

\font\tenbbb=msbm10  \font\sevenbbb=msbm7  \font\fivebbb=msbm5
\skewchar\tenbbb=127 \skewchar\sevenbbb=127 \skewchar\fivebbb=127
\newfam\bbbfam                    
\textfont\bbbfam=\tenbbb \scriptfont\bbbfam=\sevenbbb
\scriptscriptfont\bbbfam=\fivebbb
\def\Bbb{\fam\bbbfam}      

\font\tenams=msam10 \font\sevenams=msam7 
\font\fiveams=msam5 \skewchar\tenams=127
\skewchar\sevenams=127 \skewchar\fiveams=127
\newfam\amsfam                    
\textfont\amsfam=\tenams \scriptfont\amsfam=\sevenams
\scriptscriptfont\amsfam=\fiveams
\def\smc{\tensmc}                 

\mathchardef\square="2903         

\def\cite#1{{\rm[#1]}}            
\def\eq#1{{\rm(#1)}}              
\def\opname#1{\mathop{\rm#1}\nolimits} 
\def\refno#1. #2\par{\smallskip\item{\rm[#1]}#2\par} 
\def\today{\the\day\space \ifcase\month\or
           January\or February\or
           March\or April\or May\or
           June\or July\or August\or
           September\or October\or
           November\or December\fi
           \space\the\year}       

\def\declare#1. #2\par{\medskip   
      \noindent{\bf#1.}\rm
      \enspace\ignorespaces
      #2\par\smallskip}

\def\demo#1{\noindent{\it#1}.\enspace\ignorespaces} 
\def\enddemo{\ifmmode \ifinner \badenddemo \else \eqno\qed \fi
		\else \qed\par\smallskip \fi}  



\def\qed{\allowbreak\qquad\null\nobreak\hfill\hbox{$\square$}}


\def\Dl{\Delta}                   
\def\eps{\varepsilon}             




\def\FF{{\Bbb F}}                 
\def\R{{\Bbb R}}                  

\def\Hom{\opname{Hom}}            
\def\id{\opname{id}}              

\def\less{\setminus}              
\def\ox{\otimes}                  
\def\1{\'{\i}}                    
\def\2{\flat}                     
\def\3{\sharp}                    
\def\4{\natural}                  
\def\5{\diamond}                  
\def\7{\dagger}                   
\def\8{\bullet}                   
\def\.{\cdot}                     
\def\:{\colon}                    




\def\longto{\mathop{\longrightarrow}\limits} 


\def\tsum{\mathop{\textstyle\sum}\nolimits} 

\def\row#1#2#3{{#1}_{#2},\dots,{#1}_{#3}}   
\def\sepword#1{\qquad\hbox{#1}\quad} 
\def\set#1{\{\,#1\,\}}            
\def\textit#1{{\it#1\/}}          

\newcount\citenum
\def\reflist#1{\def\key##1{\advance\citenum by 1
                  \edef##1{\the\citenum}}#1} 

\reflist{\key\ChariP \key\ConnesKrHopf \key\ConnesKrTorino
\key\ConnesKrRHI \key\ConnesMHopf \key\ourbook \key\Hazewinkel
\key\KasselQGroups \key\Kreimer \key\KreimerChen \key\Manoukian
\key\Sweedler \key\Wulkenhaar \key\Zimmermann}


\centerline{\bf On the antipode of Kreimer's Hopf algebra}

\bigskip

\centerline{\smc H\'ector Figueroa}

\smallskip

\centerline{\it Department of Mathematics, Universidad de
Costa Rica, 2060 San Jos\'e, Costa Rica}

\medskip

\centerline{\smc Jos\'e M. Gracia--Bond\1a}

\smallskip

\centerline{\it Department\ of Physics, Universidad de
Costa Rica, 2060 San Jos\'e, Costa Rica}

\bigskip

{\narrower\eightrm\baselineskip=9.5pt       
\noindent
We give a new formula for the antipode of the algebra of rooted trees,
directly in terms of the bialgebra structure. The equivalence, proved
in this paper, among the three available formulae for the antipode,
reflects the equivalence among the Bogoliubov--Parasiuk--Hepp,
Zimmermann, and Dyson--Salam renormalization schemes.
\par}

\bigskip


\bigskip

\noindent {\bf 1. Introduction}

\bigskip

More than two years ago, Kreimer~\cite{\Kreimer} discovered that there
is a Hopf algebra structure encoding Zimmermann's forest
formula~\cite{\Zimmermann} in perturbative renormalization theory.
Shortly afterwards, an essential coincidence was found between Kreimer's
algebra and the Hopf algebras introduced by Connes and Moscovici in
connection with the index problem for $K$-cycles on
foliations~\cite{\ConnesMHopf}.

A unified treatment in terms of the algebra of rooted trees $H_R$ was
developed in~\cite{\ConnesKrHopf}: there to each (superficially
divergent) Feynman diagram a sum of rooted trees is assigned; the
assignment is straightforward when the diagram contains only disjoint
or nested subdivergences (as it then leads to a single tree), but it
does not work smoothly for overlapping divergences~\cite{\Wulkenhaar}.

The central role in the application of Kreimer--Connes--Moscovici
algebras is played by the \textit{antipode}. In~\cite{\ConnesKrHopf}
two equivalent definitions of the antipode in $H_R$ were given,
representing respectively ---in the framework of the algebra of rooted
trees--- the recursive Bogoliubov--Parasiuk--Hepp procedure for
renormalizing Feynman integrals with subdivergences, and Zimmermann's
forest formula which solves that recursion; that indeed they
correspond to the antipode of the Hopf algebra of rooted trees is
implied rather than proven.

Here we \textit{construct} the antipode for $H_R$, giving a new
formula for computing it in terms of the coproduct; and then we show
its equivalence to each of the formulae by Connes and Kreimer. It
turns out that this new formula corresponds to the Dyson--Salam
procedure for renormalization.

\bigskip

\noindent {\bf 2. The antipode of the Hopf algebra of rooted trees}

\bigskip

To establish the notation, we briefly recall some basic facts concerning
the antipode of a Hopf algebra
(consult~\cite{\ChariP,\ourbook,\KasselQGroups,\Sweedler} for
proofs), and then the algebra of rooted trees.

Given a unital algebra $(A,m,u)$ and a counital coalgebra
$(C,\Dl,\eps)$ over a field $\FF$, the \textit{convolution}
of two elements $f,g$ of the vector space of $\FF$-linear
maps $\Hom(C,A)$ is defined as the map $f * g \in \Hom(C,A)$
given by the composition
$$
C \longto^\Dl C \ox C \longto^{f\ox g} A \ox A \longto^m A.
$$
This product turns $\Hom(C,A)$ into a unital algebra, where the unit is
the map $u \circ \eps$.  In this paper $\FF$ is the field of real numbers
$\R$.

A bialgebra $H = (A,m,u,\Dl,\eps)$ in which the identity map $\id_H$
is invertible under convolution is a \textit{Hopf algebra}, and its
(necessarily unique) convolution inverse $S$ is called the
\textit{antipode}. The property $\id_H * S = S * \id_H = u \circ \eps$
boils down to the commutativity of the diagram:
$$
    \vcenter{\hbox{
    \xymatrix{
    H \ox H  \ar[d]_{\id\ox S} & H \ar[l]_(.4)\Dl \ar[r]^(.4)\Dl
    & H     \ox     H \ar[d]^{S\ox \id}      \\
    H \ox H \ar[r]^(.6)m  & H \ar[u]_{u\circ\eps} & H \ox H \ar[l]_(.6)m.}
    }}
    \eqno (2.1)
$$
In particular, if $\Dl(a) = \sum_j a'_j \ox  a''_j$, then
$$
    \eps(a) 1_H = u \circ \eps(a) = m \circ (\id \ox S) \circ \Dl(a)
     = \tsum_j a'_j S(a''_j),
    \eqno (2.2)
$$
and likewise $\eps(a) 1_H = \sum_j S(a'_j) a''_j$.

The antipode is always a unital algebra and a counital coalgebra
antihomomorphism. When H is either commutative or cocommutative,
$S^2 = \id$; in particular, $S$ is bijective in such cases. Another
important property is that a bialgebra morphism $\ell\: H \to H'$
between Hopf algebras is automatically compatible with the antipodes:
$\ell \circ S = S' \circ \ell$~\cite{\Hazewinkel,\Sweedler}.

A \textit{rooted tree} is a finite set of points, called vertices,
joined by oriented lines that do not intersect, so that all the vertices
have exactly one incoming line, except the root which has only outgoing
lines.  In particular, there is a unique branch that joins the root with
any other vertex.  One actually works with isomorphic classes of trees.
Two rooted trees are isomorphic if the number of vertices with given
length and fertility is the same for all possible choices of lengths and
fertilities, where the \textit{fertility} of a vertex is the number of
its outgoing lines and its \textit{length} is the number of lines that
make up the unique branch joining it to the root.

For concrete examples, it will be convenient to have a list of a few
isomorphic classes of rooted trees, say with four vertices or fewer:
$$
    \5 \quad t_1
    \quad
    
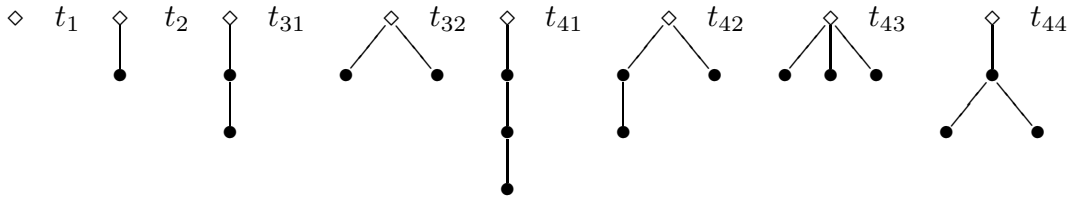
\xy     0;<18pt,0pt>: *{t_2} ;
    (0,0)*{\5};     (0,-1)*{\8}     **@{-} \endxy
    \quad
    \xy     0;<18pt,0pt>: *{t_{31}} ;
    (0,0)*{\5};     (0,-1)*{\8}     **@{-} ;
    (0,-2)*{\8}     **@{-}
    \endxy
    \quad
    \xy     0;<18pt,0pt>: *{t_{32}} ;
    (0,0)*{\5} ;
    (.8,-1)*{\8} **@{-}     , (-.8,-1)*{\8} **@{-}
    \endxy
    \quad
    \xy     0;<18pt,0pt>: *{t_{41}} ;
    (0,0)*{\5} ; (0,-1)*{\8} **@{-} ;
    (0,-2)*{\8}     **@{-} ; (0,-3)*{\8} **@{-}
    \endxy
    \quad
    \xy     0;<18pt,0pt>: *{t_{42}} ;
    (0,0)*{\5} ; (.8,-1)*{\8} **@{-}, (-.8,-1)*{\8} **@{-} ;
    (-.8,-2)*{\8}
    **@{-}
    \endxy
    \quad
    \xy     0;<18pt,0pt>: *{t_{43}} ; (0,0)*{\5} ;
    (-.8,-1)*{\8} **@{-} , (0,-1)*{\8} **@{-}, (.8,-1)*{\8} **@{-}
    \endxy
    \quad
    \xy     0;<18pt,0pt>: *{t_{44}} ;
    (0,0)*{\5} ; (0,-1)*{\8} **@{-} ;
    (-.8,-2)*{\8} **@{-} , (.8,-2)*{\8}     **@{-}
    \endxy
$$

A \textit{simple cut} $c$ of a tree $T$ is a subset of its lines such
that the path from the branch to any other vertex includes at most one
line of~$c$.  Deleting the cut branches produces several subtrees; the
component containing the original root (the trunk) is denoted $R_c(T)$.
The remaining branches also form rooted trees, where in each case the
new root is the vertex immediately below the deleted line; $P_c(T)$
denotes the set of these pruned  branches. Here, for instance, are the
possible simple cuts of~$t_{42}$:
$$
    
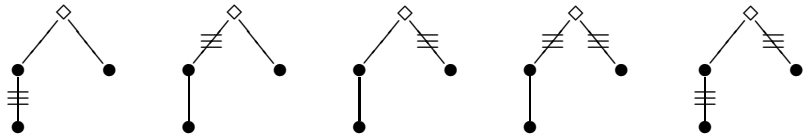
\xy     0;<18pt,0pt>:
    (0,0)*{\5} ; (.8,-1)*{\8} **@{-} ,
    (-.8,-1)*{\8}  **@{-} ; (-.8,-2)*{\8} **@{-} ?*{\equiv}
    \endxy
    \qquad
    \xy     0;<18pt,0pt>:
    (0,0)*{\5} ; (.8,-1)*{\8} **@{-} ,
    (-.8,-1)*{\8}  **@{-} ?*{\equiv} ?(1) ; (-.8,-2)*{\8}
    **@{-}
    \endxy
    \qquad
    \xy     0;<18pt,0pt>:
    (0,0)*{\5} ; (.8,-1)*{\8} **@{-} ?*{\equiv}     ?(1) ,
    (-.8,-1)*{\8}  **@{-} ; (-.8,-2)*{\8} **@{-}
    \endxy
    \qquad
    \xy     0;<18pt,0pt>:
    (0,0)*{\5} ; (.8,-1)*{\8} **@{-} ?*{\equiv}     ?(1) ,
    (-.8,-1)*{\8}  **@{-} ?*{\equiv} ?(1) ; (-.8,-2)*{\8} **@{-}
    \endxy
    \qquad
    \xy     0;<18pt,0pt>:
    (0,0)*{\5} ; (.8,-1)*{\8} **@{-} ?*{\equiv}     ?(1) ,
    (-.8,-1)*{\8}  **@{-} ; (-.8,-2)*{\8} **@{-} ?*{\equiv}
    \endxy
$$

The set of nontrivial simple cuts of a tree~$T$ will be denoted by
$C(T)$; we consider also the ``empty cut'' $c = \emptyset$, for which
$R_\emptyset(T) = T$ and $P_\emptyset(T) = \emptyset$.

The \textit{algebra of rooted trees} $H_R$ is the commutative
algebra generated by symbols $T$, one for each isomorphism class of
rooted trees, plus a unit~$1$ corresponding to the empty tree; the
product of trees is written as the juxtaposition of their symbols.
The counit $\eps: H_R \to \R$ is the linear map defined by
$\eps(1) := 1_\R$ and $\eps(T_1T_2\dots T_n) = 0$ if $\row T1n$
are trees. Kreimer defined a map $\Dl \: H_R \to H_R \ox H_R$ on the
generators, extending it as an algebra homomorphism, as follows:
$$
\Dl 1 := 1 \ox 1; \quad
\Dl T := T \ox 1 + 1 \ox T + \sum_{c\in C(T)} P_c(T) \ox R_c(T).
\eqno (2.3)
$$
Notice that $P_c(T)$ is the product of the several subtrees pruned by
the cut~$c$.  For instance,
$$
\eqalignno{
\Dl(t_1) &= t_1 \ox 1 + 1 \ox t_1,
\cr
\Dl(t_2) &= t_2 \ox 1 + 1 \ox t_2 + t_1 \ox t_1,
\cr
\Dl(t_{31}) &= t_{31} \ox 1 + 1 \ox t_{31} + t_2 \ox t_1 + t_1 \ox t_2,
\cr
\Dl(t_{32}) &= t_{32} \ox 1 + 1 \ox t_{32} + 2t_1 \ox t_2
    + t_1^2 \ox t_1,
\cr
\Dl(t_{41}) &= t_{41} \ox 1 + 1 \ox t_{41} + t_{31} \ox t_1
    + t_2 \ox t_2 + t_1 \ox t_{31},
\cr
\Dl(t_{42}) &= t_{42} \ox 1 + 1 \ox t_{42} + t_1 \ox t_{32}
 + t_2 \ox t_2 + t_1 \ox t_{31} + t_2 t_1 \ox t_1 + t_1^2 \ox t_2,
\cr
\Dl(t_{43}) &= t_{43} \ox 1 + 1 \ox t_{43} + 3t_1 \ox t_{32}
    + 3t_1^2 \ox t_2 + t_1^3 \ox t_1,
\cr
\Dl(t_{44}) &= t_{44} \ox 1 + 1 \ox t_{44} + t_{32} \ox t_1
    + 2 t_1 \ox t_{31} + t_1^2 \ox t_2.
& (2.4) \cr}
$$

A most useful tool is the sprouting of a new root; namely the
morphism $L\: H_R \to H_R$ given by the linear map defined by
$$
L(T_1 \dots T_k) := T,
$$
where $T$ is the rooted tree obtained by conjuring up a new vertex as
its root and extending lines from this vertex to each root of
$\row T1k$. For instance,
$$
    L\Bigl( \vcenter{
       \xy 0;<10pt,0pt>: (0,0)*{\5} ; (.8,-1)*{\8} **@{-},
					     (-.8,-1)*{\8} **@{-} \endxy} \Bigr)
    = \vcenter{     \xy     0;<10pt,0pt>: (0,0)*{\5}; (0,-1)*{\8} **@{-};
				    (-.8,-2)*{\8} **@{-}, (.8,-2)*{\8} **@{-}\endxy}
    \sepword{and}
    L\Bigl( \vcenter{\xy 0;<10pt,0pt>: (0,0)*{\5};
					      (0,-1)*{\8} **@{-} \endxy} \
		     \vcenter{\xy 0;<10pt,0pt>:     (0,0)*{\5};
					      (0,-1)*{\8} **@{-} \endxy} \Bigr)
    = \vcenter{     \xy     0;<10pt,0pt>: (0,0)*{\5} ;
				    (-.8,-1)*{\8} ="2" **@{-}, (.8,-1)*{\8} **@{-};
				    (.8,-2)*{\8} **@{-}; (-.8,-2)*{\8};     "2"     **@{-}
			    \endxy}
    \eqno (2.5)
$$

The proof that $\Dl$ is indeed a coproduct is based on the formula
$$
\Dl \circ L = L \ox 1 + (\id \ox L) \circ \Dl.
\eqno (2.6)
$$
For details see~\cite{\ConnesKrHopf} or~\cite{\ourbook}.

When dealing with particular Hopf algebras, the antipode is often
determined by specific properties of the algebras in question, and the
defining property of the antipode is scarcely used.  The latter turns
out to be extremely useful in our context, however.  We compute the
antipode $S\: H_R \to H_R$ by exploiting its very definition as the
convolution inverse of the identity in $H_R$, via a geometric series:
$$
S := (\id)^{*-1} = (u \circ \eps -(u \circ \eps -\id))^{*-1}
  = u \circ \eps + (u \circ \eps -\id) + (u \circ \eps -\id)^{*2}
    + \cdots
$$

\proclaim Lemma 2.1.
If $T$ is a rooted tree with $n$ vertices, the geometric series
expansion of $S(T)$ has at most $n + 1$ terms.

\demo{Proof}
The claim is certainly true for $t_1$. Assume that it holds for all
trees with $n$ vertices. Let $T$ be a rooted tree with $n + 1$
vertices; then
$$
\eqalign{
(u \circ \eps -\id)^{*(n+2)}(T)
&= (u \circ \eps -\id) * (u \circ \eps -\id)^{*(n+1)}(T)
\cr
&= m \circ [(u \circ \eps -\id) \ox (u \circ \eps -\id)^{*(n+1)}]
     \circ \Dl(T)
\cr
&= m \circ [(u \circ \eps -\id) \ox (u \circ \eps -\id)^{*(n+1)}]
\cr
&\qquad\qquad \biggl( T \ox 1 + 1 \ox T
                      + \sum_{c\in C(T)} P_c(T) \ox R_c(T) \biggr).
\cr}
$$
The first and second term vanish because $(u \circ \eps - \id)1 = 0$.
By the induction hypothesis the third term is zero.
\enddemo

As an immediate corollary we obtain that $S$ so defined is indeed the
antipode.

One of the advantages of this formulation is that we obtain a fully
explicit formula for $S$ from the coproduct table.  If $a \in H^n$,
$\Dl(a) = \sum_{i_1} a'_{i_1} \ox a''_{i_1}$, $\Dl(a''_{i_1}) =
\sum_{i_2} a'_{i_1i_2} \ox a''_{i_1i_2}$ and in general $\Dl(a''_{\row
i1k}) = \sum_{i_{k+1}} a'_{\row i1{k+1}} \ox a''_{\row i1{k+1}}$, then
$$
(u \circ \eps -\id)^{*k+1}(a) = (-1)^{k+1} \sum_{\row i1k}
b'_{i_1} b'_{i_1 i_2} \cdots b'_{\row i1k} b''_{\row i1k},
$$
where
$$
\eqalignno{
b'_{\row i1j} &:= \cases{
  0  & if $a'_{\row i1j} = 1$ or $a''_{\row i1j} = 1$, \cr
  a'_{\row i1j} & otherwise, \cr}
\cr
\noalign{\hbox{and}}
b''_{\row i1j} &:= \cases{ 0  & if $a''_{\row i1j} = 1$, \cr
                          a''_{\row i1j} & otherwise. \cr}
\cr}
$$
For instance, using~\eq{2.4},
$$
\eqalignno{
S(t_{42})
&= - t_{42} + (t_1t_{32} + t_2^2 + t_1t_{31} + 2 t_1^2t_2)
   - (5 t_1^2t_2 + 2 t_1^4) + 3 t_1^4
\cr
&= - t_{42} + t_1 t_{32} + t_2^2 + t_1 t_{31} - 3t_1^2 t_2  + t_1^4.
& (2.7) \cr}
$$
Similarly, if we denote by $t'$ the rooted tree in~\eq{2.5} with 5
vertices, then
$$
\eqalignno{
S(t') &= - t' + (2 t_1t_{42} + 2 t_2t_{31} +  t_1^2t_{32} + 3 t_1t_2^2)
\cr
&\qquad
-(2 t_1^2t_{32} +6 t_1t_2^2 + 2 t_1^2t_{31} + 8 t_1^3t_2 + t_1^5)
+ (12 t_1^3 t_2 + 6 t_1^5) - 6 t_1^5
\cr
&= - t' + 2 t_1t_{42} + 2 t_2t_{31} - t_1^2t_{32} - 3 t_1t_2^2
- 2 t_1^2t_{31} + 4 t_1^3t_2 - t_1^5.
\cr}
$$

The reader will find without difficulty that these formulae correspond
to the original Dyson--Salam procedure for renormalizing Feynman
graphs with subdivergences: see, for instance,~\cite{\Manoukian}.

In correspondence with Bogoliubov's recursive formula for
renormalization, equations $m \circ (S \ox \id) \circ \Dl(T) = 0$
and~\eq{2.4} suggest to define the antipode recursively,
as indeed done by Connes and Kreimer:
$$
S_B(T) := - T - \sum_{c\in C(T)} S_B(P_c(T)) R_c(T).
$$
For instance,
$$
S_B(t_{42}) = -t_{42} - S_B(t_1) t_{32} - S_B(t_2) t_2
- S_B(t_1) t_{31}  - S_B(t_2t_1) t_1 - S_B(t_1^2) t_2,
$$
which gives again~\eq{2.7}.  We next check  that $S_B$ is indeed the
antipode.

\proclaim Proposition 2.2.
If $T$ is any rooted tree, then $S(T) = S_B(T)$.

\demo{Proof}
For convenience, we abbreviate $\eta := u\circ\eps - \id$.  The
statement holds, by a direct check, if $T$ has $1,2$ or~$3$ vertices.
If it holds for all rooted trees with at most $n$ vertices and if $T$ is
a rooted tree with $n+1$ vertices, then
$$
\eqalignno{
S(T) &= \eta(T) + \sum_{j=1}^n \eta^{*j} * \eta(T)
 = -T + m \circ \biggl( \sum_{j=1}^n \eta^{*j} \ox \eta \biggr)
   \circ \Dl(T)
\cr
&= -T + m \circ \sum_{j=1}^n \eta^{*j} \ox \eta
   \biggl(T \ox 1 + 1\ox T + \sum_{c\in C(T)} P_c(T) \ox R_c(T)\biggr)
\cr
&= -T - \sum_{c\in C(T)} \sum_{j=1}^n \eta^{*j} (P_c(T)) \, R_c(T)
\cr
&=  -T - \sum_{c\in C(T)} S_B(P_c(T)) \,R_c(T) = S_B(T),
\cr}
$$
where the penultimate equality uses the inductive hypothesis.
\enddemo

Zimmermann's forest formula corresponds to the following
nonrecursive formula for the antipode:
$$
S_Z(1) := 1,\quad S_Z(T) := - \sum_{d\in D(T)} (-1)^{\#d} P_d(T) R_d(T),
$$
where $D(T)$ is the set of all cuts, not necessarily simple, including
the empty cut, and $\#d$ is the cardinality of~$d$.

\proclaim Proposition 2.3.
If $T$ is any rooted tree, then $S(T) = S_Z(T)$.

\demo{Proof}
First we prove that, for an arbitrary rooted tree $T$,
$$
S(L(T)) = - L(T) - S(T)\,t_1 - \sum_{c\in C(T)} S(P_c(T))\,L(R_c(T)).
\eqno (2.8)
$$
Indeed, if $T$ has $n$ vertices, then, by Lemma~2.1 and~\eq{2.6},
$$
\eqalignno{
S(L(T))
&= - L(T) + m \circ \biggl( \sum_{j=1}^n \eta^{*j} \ox \eta \biggr)
     \circ \Dl(L(T))
\cr
&= - L(T) + m \circ \sum_{j=1}^n \eta^{*j} \ox \eta
      \bigl( L(T) \ox 1 + (\id \ox L) \circ \Dl(T) \bigr)
\cr
&= - L(T) + m \circ \sum_{j=1}^n \eta^{*j} \ox \eta
\cr
&\qquad\qquad \biggr( L(T) \ox 1 + T \ox t_1  + 1 \ox L(T)
        + \sum_{c\in C(T)} P_c(T) \ox L(R_c(T)) \biggl)
\cr
&= - L(T) - S(T)\,t_1 - \sum_{c\in C(T)} S(P_c(T))\,L(R_c(T)).
\cr}
$$

Suppose that $S_Z$ were also to satisfy~\eq{2.8}.  Since any rooted tree
can be written as an image of $L$, and on the right side $S$ is applied
only to rooted trees of strictly fewer vertices, the proposition will
follow by induction on the number of the vertices.  It remains,
therefore, to prove that~\eq{2.8} holds for~$S_Z$.

For a given rooted tree $T$, let $\ell_0$ be the new line in $L(T)$,
and $\row v1k$ the vertices of length one with respect to the root of
$T$. Now $D(L(T)) = A \uplus B$ where $d \in A$ or~$B$ according as
$\ell_0 \in d$ or not. Thus,
$$
S_Z(L(T)) = - \biggl( \sum_{d\in A} + \sum_{d\in B} \biggr)
             (-1)^{\#d} P_d(L(T))\,R_d(L(T)).
\eqno (2.9)
$$
If $d \in A$, then $e = d \less \{\ell_0\}$ is a cut of $T$; moreover,
$R_d(L(T)) = t_1$, $P_d(L(T)) = P_e(T) R_e(T)$ and $\#d = \#e + 1$, so
that the first sum of~\eq{2.9} equals
$-S(T)\,t_1$.

For a given $d \in B \less \{\emptyset\}$ and each $j \in K =:
\set{1,\dots,k}$, let $\ell_j$ be the line in~$d$ closer to the root
that is linked to $v_j$ (if any).  Then $c' := \set{\ell_j \: j \in K}$
is a simple cut of $T$.  If $\#c'$ is odd, we set $c := c'$, whereas if
$\#c'$ is even, we set $c := c' \less \{\ell_s\}$, where $s$ is the
smallest integer in $K$ for which there is a line with the required
property.  In either case, we take $e := d \less c$.  Clearly $R_d(L(T))
= L(R_c(T))$, $(-1)^{\#d+1} = (-1)^{\#e}$, and $P_d(L(T)) = P_e(T_c)
R_e(T_c)$, where we use the temporary notation $T_c := P_c(T)$.  It
follows that the second sum of~\eq{2.9} equals
$$
\sum_{c\in C(T)} \sum_{e\in D(T_c)} (-1)^{\#e}
  P_e(T_c)\, R_e(T_c)\, L(R_c(T))
 = - \sum_{c\in C(T)}S_Z(P_c(T))\, L(R_c(T)).
$$
Finally, since the summand for the empty cut is $-L(T)$, the
proposition is proved.
\enddemo

In summary, modulo the distiction between the antipode and the
``twisted'' or ``renormalized''
antipode~\cite{\ConnesKrTorino,\ConnesKrRHI,\KreimerChen}, Kreimer's
algebraic approach reflects, within the framework of the algebra of
rooted trees, the equivalence of the Dyson--Salam, the
Bogoliubov--Parasiuk--Hepp and the Zimmermann procedures for
renormalizing Feynman diagrams.

\bigskip\bigskip

\noindent {\bf Acknowledgements}

\smallskip

The geometric series trick was pointed out to us by Earl Taft; we are
greatly indebted to him. We thank Stephen A. Fulling for a good
suggestion and Joseph C. V\'arilly for illuminating discussions and
\TeX{}nical help. We acknowledge support from the Vicerrector\1a de
Investigaci\'on de la Universidad de Costa Rica.

\bigskip

\noindent {\bf References}

\frenchspacing

\bigskip

\refno\ChariP.
V. Chari and A. Pressley,
\textit{A Guide to Quantum Groups},
Cambridge University Press, Cambridge, 1994.

\refno\ConnesKrHopf.
A. Connes and D. Kreimer,
``Hopf algebras, renormalization and noncommutative geometry'',
Commun. Math. Phys. {\bf 199} (1998), 203--242.

\refno\ConnesKrTorino.
A. Connes and D. Kreimer,
``Renormalization in quantum field theory and the Rie\-mann--Hilbert
problem'',
hep-th/9909126, IHES, Bures-sur-Yvette, 1999.

\refno\ConnesKrRHI.
A. Connes and D. Kreimer,
``Renormalization in quantum field theory and the Rie\-mann--Hilbert
problem I: the Hopf algebra structure of graphs and the main
theorem'',
hep-th/9912092, IHES, Bures-sur-Yvette, 1999.

\refno\ConnesMHopf.
A. Connes and H. Moscovici,
``Hopf algebras, cyclic cohomology and the transverse index theorem'',
Commun. Math. Phys. {\bf 198} (1998), 198--246.

\refno\ourbook.
J. M. Gracia--Bond\1a, J. C. V\'arilly and H. Figueroa,
\textit{Elements of Noncommutative Geo\-metry}, Birkh\"auser, Boston,
2000, forthcoming.

\refno\Hazewinkel.
M. Hazewinkel,
\textit{Formal Groups and Applications},
Academic Press, New York, 1978.

\refno\KasselQGroups.
C. Kassel,
\textit{Quantum Groups},
Springer, Berlin, 1995.

\refno\Kreimer.
D. Kreimer,
``On the Hopf algebra structure of perturbative quantum field
theories'',
Adv. Theor. Math. Phys. {\bf 2} (1998), 303--334.

\refno\KreimerChen.
D. Kreimer,
``Chen's iterated integral represents the operator product
expansion'',
Adv. Theor. Math. Phys. {\bf 3} (1999), 3;
hep-th/9901099.

\refno\Manoukian.
E. B. Manoukian,
\textit{Renormalization},
Academic Press, London, 1983.

\refno\Sweedler.
M. E. Sweedler,
\textit{Hopf algebras},
Benjamin, New York, 1969.

\refno\Wulkenhaar.
R. Wulkenhaar,
``On Feynman graphs as elements of a Hopf algebra'',
hep-th/9912220, Luminy, 1999.

\refno\Zimmermann.
W. Zimmermann,
``Convergence of Bogoliubov's method of renormalization in momentum
space'',
Commun. Math. Phys. {\bf 15} (1969), 208--234.

\bye